\documentclass[final,3p,times,twocolumn]{elsarticle}
\usepackage{amssymb}
\usepackage{amsmath}
\usepackage{mathtools}
\usepackage{hyperref}
\usepackage{booktabs}
\usepackage{color}
\usepackage{tabularx}
\usepackage{algorithm}  
\usepackage{algpseudocode}

\newcounter{bla}

\journal{Computer Physics Communications}

\begin{document}

\begin{frontmatter}



\title{PairDiag: an exact diagonalization program for solving general pairing Hamiltonians}

\author[a,b,c]{Xiao-Yu Liu}
\author[a]{Chong Qi\corref{author}}

\cortext[author]{Corresponding author.\\\textit{E-mail address:} chongq@kth.se}
\address[a]{Department of Physics, Royal Institute of Technology, Stockholm 104 05, Sweden}
\address[b]{Institute of Modern Physics, Chinese Academy of Sciences, Lanzhou 730000, China}
\address[c]{University of Chinese Academy of Sciences, Beijing 100049, China}

\begin{abstract}

We present a program for solving exactly the general pairing Hamiltonian based on diagonalization. The program generates the seniority-zero shell-model-like basis vectors via the `01' inversion algorithm. The Hamiltonian matrix is constructed in that seniority-zero space. The program evaluates all non-zero elements of the Hamiltonian matrix  ``on the fly" using the scattering operator and the search algorithm that act on the generated basis. The matrix is diagonalized by using the iterative Lanczos algorithm. The program thus developed, PairDiag, can  calculate efficiently the ground-state eigenvalue and eigenvector of any pairing Hamiltonian. The program can be easily implemented to replace the BCS approximation in standard self-consistent mean-field calculations. The code is parallelized using OpenMP. For larger systems with dimension around 10$^{8-9}$, the calculation can be done within a day on standard desktop computers.

\end{abstract}


\end{frontmatter}

\begin{small}
	\noindent
	{\bf PROGRAM SUMMARY}\\
	{\em Program Title:} PairDiag.\\                                      
	{\em Licensing provisions:} CC by NC 3.0\\
	{\em Programming language:} Fortran 90.\\
	{\em Computer:} All computers with a Fortran compiler supporting at least Fortran 90.\\                         
	{\em Operating system:} Linux.\\
	{\em RAM:} RAM needed depends on the dimension of calculation.\\
	{\em Number of processors used:} No built-in limit under OpenMP\\
	{\em Nature of problem:} The standard pairing problem can be solved directly by diagonalizing the Hamiltonian matrix.\\
	{\em Solution method:} This program solves the lowest eigenvalue and corresponding eigenvector in a shell model style by the restart reorthogonalization Lanczos plus QR method. The pairing Hamiltonian matrix is constructed in the seniority-zero space.\\
	{\em Restrictions:} The total number of orbits involved must be less than 63, and the dimension should also be limited to the order of 10$^{8-9}$ for efficient diagonalization on a standard desktop.\\

\end{small}

\section{Introduction}

It is well established that the pairing correlation is an essential ingredient in describing the ground-state properties of finite atomic nuclei, and the solving of the pairing Hamiltonian is important for describing not only nuclei but also other many-body systems~\cite{duke2004} including neutron stars~\cite{dean2003}, superconductors~\cite{brog2013}, and trapped two-component Fermi gases~\cite{ring1980}. The nuclear pairing theory based on the Bardeen-Cooper-Schrieffer (BCS) approximation~\cite{bard1957} was considered for the first time 60 years ago~\cite{bohr1958}. But the BCS condensate with indefinite particle numbers is problematic in weak pairing finite systems, and the pairing condensate around shell closures will collapse. To overcome those drawbacks, there have been extensive efforts in developing particle-number-conserved approximations (see, Ref.~\cite{jiax2019} and references therein) and exact pairing models which include the diagonalization~\cite{voly2001, chen2014, xuuu2013, chan2015, moli1997, zeng1983} and the Richardson (or the Richardson-Gaudin) method~\cite{rich1963, guan2012, guan2014}. In 1966, Richardson~\cite{rich1966} studied the exact and the BCS solutions of the standard pairing Hamiltonian with $\Omega = 8\sim 32$ at half-fillings, and concluded that the BCS model strongly underestimates the pairing correlations even for relatively large pairing strength. 

The Richardson algebraic approach can be applied for large systems but is limited to the standard pairing Hamiltonians with constant pairing force. Therefore, it is important to develop algorithms that can handle the general pairing Hamiltonians. The most straightforward method for the numerically exact solution is to diagonalize the Hamiltonian in configuration spaces of fixed seniority~\cite{chen2014, moli1997}. In diagonalization, the number of particles and orbits that can be included is usually limited due to the rapid growth of the dimension. In our previous calculations with exact diagonalization~\cite{xuuu2013, chan2015}, the dimension of the problem is limited to 10$^{6}$ on standard desktops. The limitation is related to the inefficient generation and storage of the basis vectors and the high time complexity of the matrix operations. The limited capacity of the diagonalization makes it challenging to perform realistic calculations and to compare the results with those of BCS or other approximate approaches.

In this study, we developed an efficient diagonalization program, PairDiag, for solving the general pairing Hamiltonian in the doubly-degenerate deformed system. We have applied in the code a novel basis generation algorithm, dynamic evaluation of the non-zero Hamiltonian matrix elements, and the Lanczos~\cite{lanc1950} algorithm for diagonalization. The system with odd-number of particles can also be treated within the blocking approximation. The program is optimized using OpenMP parallelization and packaged in a fortran module which allows it to be easily combined with existing nuclear structure programs (e.g., self-consistent mean field codes EV8~\cite{bonc2005} and Sky3D~\cite{maru2018}) as an alternate to the problematic BCS solver. The vector generation and search algorithms in the program can also be transplanted to other programs including the large-scale shell model code.

\section{The General Pairing Hamiltonian}

The general pairing Hamiltonian acting in doubly-degenerate time-reversed states is given by
\begin{equation}
\label{hamiltonian1}
\hat{H} = \sum_{i}\epsilon_{i}a_{i}^{\dagger}a_{i}^{}+\sum_{ii'}V_{ii'}a_{i}^{\dagger}a_{\bar{i}}^{\dagger}a_{\bar{i'}}^{}a_{i'}^{}
\end{equation}
\noindent where $\epsilon_{i}$ are the single-particle energies and $V_{ii'}$ are the orbit-dependent pairing interaction strength. $a_{i}^{\dagger}$ and $a_{i}^{}$ is the particle creation and annihilation operator, respectively. The Hamiltonian with $V_{ii'}=G$, a constant strength, is usually called the standard pairing Hamiltonian. For the convenience of description, we will use the orbit $i$ to refer to the time-reversed states ${i}$ and $\bar{i}$, and treat two particles distributed in the time-reversed states as one pair. We can then define $S^{+}_{i}=a_{i}^{\dagger}a_{\bar{i}}^{\dagger}$ and $S^{-}_{i}=a_{\bar{i}}^{}a_{i}^{}$ as the pair creation and annihilation operator of orbit $i$. For a system of even-number paired particles in the finite space with $\Omega$ orbits, the Hamiltonian can be written as 
\begin{equation}
\label{hamiltonian2}
\hat{H} = \sum_{i}^{\Omega}(2\epsilon_{i}+V_{ii})S^{+}_{i}S^{-}_{i}+\sum_{i\neq j}^{\Omega}V_{ij}S^{+}_{i}S^{-}_{j}
\end{equation}
\noindent $S^{+}_{i}S^{-}_{i}$ and $S^{+}_{i}S^{-}_{j}$ can be understood as the pair number and the scattering operator. $V_{ii}$ is the diagonal pairing element which is sometimes referred to the self-energy of a pair. For an odd-mass system with only one unpaired particle, the odd particle blocks the pairs scattering into the orbit occupied by itself due to the Pauli principle (blocking effect). The space can be expressed as the direct sum of seniority-one subspaces corresponding to different blocked orbits, and the Hamiltonian can be given by 
\begin{equation}
\label{hamiltonian3}
\hat{H}_{b} = \epsilon_{b}a_{b}^{\dagger}a_{b}^{}+\sum_{i\neq b}^{\Omega}(2\epsilon_{i}+V_{ii})S^{+}_{i}S^{-}_{i}+\sum_{i\neq j\neq b}^{\Omega}V_{ij}S^{+}_{i}S^{-}_{j}
\end{equation}
\noindent where $\hat{H}_{b}$ is the Hamiltonian for the subspace in which orbit $b$ is blocked.

\section{Principles of the Method}

In the present program, we solve the pairing Hamiltonian via diagonalization to get the ground-state eigenvalue and eigenvector. This shell-model style approach can be divided into three parts, first generating the basis with fixed seniority, then constructing and diagonalizing the Hamiltonian matrix. In the following content, we focus on the Hamiltonian in Eq.~\ref{hamiltonian2} for the even-mass seniority-zero system, while the Hamiltonian in Eq.~\ref{hamiltonian3} for the odd-mass seniority-one system can be solved based on the even-mass system.

\subsection{Basis Generation}

Let's consider an even-mass system with $m$ pairs and $n$ orbits ($m\le n$). In the seniority-zero scheme, the basis consists of all possible Slater determinants of $m$ identical pairs distributed in $n$ different orbits. Each determinant can be represented by a binary word in the computer, while each bit of the word being associated to an individual orbit, with a value of 1 or 0 depending on whether the orbit is fully occupied. A set of all binary numbers with $m$ occupied bits distributed in the first $n$ digits is equivalent to a seniority-zero space, and the dimension of the space from the binomial coefficient is
\begin{equation}
\label{extcmn}
C^m_n = 
\left\{
\begin{aligned}
&\frac{n!}{m!(n-m)!} &n\ge m \\
&        0           &n<m                   
\end{aligned}
\right.
\end{equation} 
\noindent The definition of extension to $n<m$ will be used in the vector hash search. For the case where 2 pairs occupy 4 orbits, a set of 6 binary numbers from 0011B to 1100B can be used to represent the space.

To generate all the required binary numbers efficiently, an iterative combination algorithm, `01' inversion algorithm, is used. Each iteration takes an integer as input and searches from its lowest binary digit until the 2 adjacent bits with pattern `01' is found, then the found `01' will be inverted to `10', and all bits `1' below the turned `10' will be moved to refill this number from the lowest digit. After the two steps, a larger integer is obtained which will be the input for the next iteration. The pseudocode shown in Algorithm.~\ref{conversion} is an efficient implementation based on bit operations. Since one iteration only calculates one next larger integer while keeping the total number of occupied bits conserved, the minimum and the maximum integer in the space representing the start and the end of the iteration must be specified in advance, and the remaining numbers will be calculated iteratively from the minimum.

\begin{table}
\small
\begin{center}
\setlength{\tabcolsep}{6.1mm}{
\caption{\label{data} Index and binary values of all integers in the space of 2 pairs in 4 orbits, The decimal values shown displays the ascending order.}
\begin{tabular}{ccc}
\hline
Index &  Binary value &  Decimal value\\
\hline
1     &  0011         &  03           \\
2     &  0101         &  05           \\
3     &  0110         &  06           \\
4     &  1001         &  09           \\
5     &  1010         &  10           \\
6     &  1100         &  12           \\
\hline
\end{tabular}}
\end{center}
\end{table}

\begin{algorithm} 
\caption{\label{conversion} `01' inversion algorithm. BTEST(), IBCLR(), and others refer to the Fortran intrinsic bit manipulation functions} 
\begin{algorithmic}
\Require integer I$_{in}$
\Ensure integer I$_{out}$
\State I$_{tail}$ = 0                                                 
\For{i = 0, 2, $\cdots$, 61} 
\If {(BTEST(I$_{in}$, i))}  
\State I$_{in}$ = IBCLR(I$_{in}$, i)
\If {(!BTEST(I$_{in}$, i+1))}  
\State I$_{in}$ = IBSET(I$_{in}$, i+1)
\State \textbf{exit}  
\EndIf
\State I$_{tail}$ = IBSET(ISHFT(I$_{tail}$, 1), 0)
\EndIf 
\EndFor
\State I$_{out}$ = I$_{in}$ + I$_{tail}$
\State \textbf{return} I$_{out}$
\end{algorithmic}
\end{algorithm}

For the previous example of $C_{4}^{2}$ with the minimum 0011B and the maximum 1100B, the iteration should start at 0011B and end when the output reaches 1100B. In the first iteration, the `01' in the 2nd and 3rd digits of 0011B should be inverted to `10' to get 0101B, because the bit `1' in the 1st digit is already at the lowermost, the output is 0101B. In the same way, 0110B is the second output. For the input 0110B, 1010B will be obtained by inverting the `01' and then the bit `1' in the 2nd digit needs to be moved to the lowermost to get the 1001B. Iteratively, 1010B and 1100B will be created in order, then the calculation should be terminated since 1100B has reached the maximum of the space. With five iterations starting from 0011B, all six numbers summarized in Table~\ref{data} are obtained, and the index are assigned according to the order of generation, which is also the order of the values in the space.

For a space of dimension $n$, the time complexity of the `01' inversion algorithm over the entire space can be roughly estimated as a linear order $O(n)$. In PairDiag code, a 64-bit integer is used to represent a valence vector, so the total number of orbits allowed is less than 63 after excluding the sign bit. In the calculation, all the integers created are stored in an 1D array. Each element in the array, like in Table~\ref{data}, has two properties, one is the binary value representing the wave function $|i\rangle$, and the other is the index number representing the position $i$. Since the array generated by the algorithm is strictly in ascending order and obeys a special combination rule, in addition to extracting the wave function $|i\rangle$ directly from the given index $i$, the index $i$ of any element can also be calculated from its binary value $|i\rangle$.

\subsection{Vector Search}

For locating the index $i$ for a element $|i\rangle$ in the basis array, the program provides two search methods for different situations: the binary search and the hash search. The binary search can always be used for a sorted array. The search process starts from the middle element of the array. If the middle element is exactly the element to be found, the search process ends, if the element is greater than (or less than) the middle element, the search is performed in the half of the array that is greater than (or less than) the middle element, and starts with the ``sub middle'' element as before. This search algorithm makes full use of the order relationship between the elements in a divide and conquer strategy by halving the search range after every comparison. The search task can be completed within $O($log$\,n)$ in the worst case. The use of binary search in the program is when the total number of base vectors is less than $C_{n}^{m}$ due to the truncation, which will be introduced later.

Without truncation, the total element number is equal to $C_{n}^{m}$, and the structure of array remains intact, a more efficient hash search from a function $f(|i\rangle)=i$ is used in the program. To write the hash function, we use $O_{m}$ to represent the orbit number for the $m\,$th occupied orbit in the vector $|i\rangle$. For example 11010B, there are three occupied bits, $O_{1}$ for the first occupied bit is 2, $O_{2}$ for the second one is 4, $O_{3}=5$ is for the third one. With these definitions, we can represent the function $f(|i\rangle)$ for a vector $|i\rangle$ with $m$ occupied bits as 
\begin{equation}\nonumber
\label{hash}
f(|i\rangle) = 1+\sum_{j}^{m}C^{j}_{O_{j}-1}
\end{equation} 
\noindent In Table~\ref{data}, $f(0011B)=1+C^{1}_{0}+C^{2}_{1}=1$, $f(1001B)=1+C^{1}_{0}+C^{2}_{3}=4$, and $f(1100B)=1+C^{1}_{2}+C^{2}_{3}=6$. The time complexity of this hash search algorithm is $O(1)$.

\subsection{Matrix Construction}

With the basis generated and the search algorithms provided, we can now construct the sparse Hamiltonian matrix in an efficient way by evaluating all the non-zero matrix elements directly. The diagonal elements in the Hamiltonian matrix are usually non-zero, and the value of elements $H_{i, i}=\langle i|\hat{H}|i\rangle$ determined by the first item of the Hamiltonian in Eq.~\ref{hamiltonian2} is
\begin{equation}
\label{diagonal}
H_{i, i} = \sum_{n}^{\Omega}(2\epsilon_{n}+V_{nn})\langle i|S^{+}_{n}S^{-}_{n}|i\rangle
\end{equation}
\noindent Only when the orbit $n$ in vector $|i\rangle$ is fully occupied, the value of $\langle i|S^{+}_{n}S^{-}_{n}|i \rangle$ will be 1, otherwise 0.

The value of non-diagonal elements $H_{i, j}=H_{j, i}$ is determined by the second term of the Eq.~\ref{hamiltonian2}. For a vector $|i \rangle$ with index $i$ in a space of $C_{n}^{m}$, if we mark one of the $m$ occupied orbits as $O$ and one of the $n$-$m$ empties as $E$, then ``scatter'' the pair from $O$ to $E$ with the pair scattering operator to form a new vector $|j \rangle=S^{+}_{E}S^{-}_{O}|i\rangle$, the matrix element $\langle i|\hat{H}|j\rangle=V_{EO}$ will be non-zero if $V_{EO}\neq 0$. The position of this element $(i, j)$ in the matrix can be obtained by searching the index $j$ of the vector $|j\rangle$. Combining the different $O$ and $E$ in $|i \rangle$, the total number of such $|j\rangle$ and also the non-zero $H_{i, j}$ is $m(n-m)$. In the PairDiag code, binary search or hash search is used to locate index of different $|j\rangle$. Through the simple bit operations and search, we can evaluate all the non-zero elements in the Hamiltonian matrix directly. 

Still using the previous example in Table~\ref{data} with assigning single-particle energies from 1 to 4, and using the constant -0.2 as the overall pairing interaction strength. The Hamiltonian can be expressed as a 6$\times$6 real symmetric matrix. For the first row, the diagonal element referring to Eq.~\ref{diagonal} is $H_{1, 1} = \langle 0011|\hat{H}|0011\rangle$ = (2$\times$1-0.2)+(2$\times2$-0.2) = 5.6. The position of the 4 non-diagonal elements with the element values -0.2 are (1, 3) for $|0110\rangle=S^{+}_{3}S^{-}_{1}|0011\rangle$, (1, 5) for $|1010\rangle=S^{+}_{4}S^{-}_{1}|0011\rangle$, (1, 2) for $|0101\rangle=S^{+}_{3}S^{-}_{2}|0011\rangle$ and (1, 4) for $|1001\rangle=S^{+}_{4}S^{-}_{2}|0011\rangle$. The final matrix is 
$$H = 
\begin{bmatrix*}[r]
 5.6& -0.2& -0.2& -0.2& -0.2&  0.0\\
-0.2&  7.6& -0.2& -0.2&  0.0& -0.2\\
-0.2& -0.2&  9.6&  0.0& -0.2& -0.2\\
-0.2& -0.2&  0.0&  9.6& -0.2& -0.2\\
-0.2&  0.0& -0.2& -0.2& 11.6& -0.2\\
 0.0& -0.2& -0.2& -0.2& -0.2& 13.6\\
\end{bmatrix*}.$$

\subsection{Matrix Diagonalization}

For diagonalization, the Lanczos~\cite{lanc1950} algorithm appears as the most suitable method since only the first few states of pairing Hamiltonian are needed. As a simplification of Arnoldi method~\cite{saad2011} for the Hermitian matrix, the principle of the Lanczos (in Algorithm.~\ref{lanczos}) is a projection technique on a Krylov subspace~\cite{saad2011}. From a starting vector $p$, a new vector $q_{i}$ is generated in each iteration, and these Lanczos vectors are needed when performing reorthogonalization and Rayleigh-Ritz projection. Usually high-quality results require a large number of iterations, but the computer memory to store the vectors also grows as the iteration number increases. A restart method can avoid the difficulty by limiting the maximum number of iterations, and when reaching the maximum, the process is restarted with new starting vectors. Since Algorithm.~\ref{lanczos} can start with only one vector, the most straightforward way is to use the ground-state Ritz vector.

\begin{algorithm} 
\caption{\label{lanczos} Lanczos iteration.} 
\begin{algorithmic}
\Require starting vector $p$
\Ensure Lanczos vectors $q_{i}$, tridiagonal matrix $L$
\For{i = 1, 2, $\cdots$,} 
\State $q_{i} = p/||p||$
\State $p = Hq_{i}$
\State $L$(i, i) = $\alpha_{i} = q_{i}^{T}p$ 
\State $p = p - \alpha_{i}q_{i} - \beta_{i-1}q_{i-1}$  
\State $L$(i, i+1) = $L$(i+1, i) = $\beta_{i} = ||p||$
\EndFor
\State \textbf{return} $q_{i=1, 2, \cdots,}$, $L$ 
\end{algorithmic}
\end{algorithm}

In the PairDiag program, the Lanczos~\cite{Wuxx2000} + QR~\cite{saad2011} algorithm is performed. The default starting vector is $[1,0,\cdots,0]^{T}$. During iterations, reorthogonalization to all Lanczos vectors through the Gram-Schmidt procedure is used to cure the loss of orthogonality. The maximum number of iterations is an adjustable parameter Lanc\_\,Limit. When reaching the maximum, The QR~\cite{saad2011} algorithm is performed to calculate the Ritz pairs. According to the user's choice, the program can return the ground state of this subspace without restart. and can also perform Restart Lanczos, in which the calculation will be restarted by the ground-state Ritz vector. The Hamiltonian matrix is mainly used for matrix-vector multiplication in the Lanczos iterations, all the non-zero matrix elements are calculated dynamically without taking up much memory, and this on the fly approach also reduces the time complexity since the matrix is sparse.

\subsection{Truncation}

One will need to truncate the model space if the dimension becomes too large to be handled efficiently on a desktop. Several algorithms can be considered to help truncate the model space including the so-called importance truncation approach (see, e.g.,\cite{Jiao2014}). In the present program we have implemented a simple truncation algorithm by ordering all basis vectors accordingly the values of the corresponding diagonal matrix elements and exclude all the basis vectors with diagonal matrix elements above certain cutoff factor. The users only need to define a maximum dimension to be considered, Dimension\_\,Limit. If the full-space dimension $C_{n}^{m}$ exceeds the value of Dimension\_\,Limit, the program will truncate the space to the desired number by excluding vectors with the highest values of the diagonal matrix elements. When truncation is required, the program will calculate the maximum and minimum diagonal elements of the Hamiltonian matrix, and fill all the diagonal elements into the histogram form the minimum to the maximum with 10$^{8}$ bins. A cutoff truncation value is obtained by counting the bases bottom up until the number of diagonal elements below this value is approximately equal to Dimension\_\,Limit. Finally, all vectors with diagonal elements greater than this cutoff value will be excluded. After implementing truncation, the vector space is no longer complete and the vector can only be located by the inefficient binary search.

\section{Description of the Code}

The PairDiag code is written in Fortran 95 and packaged in a Fortran module called PairDiag. The use of the module requires three steps.

\subsection{step 1: Initialize the Inputs}

4 public variables represent the inputs of the module must be explicitly initialized by the user before the calculation. The first two determine the space, and the last two define the Hamiltonian matrix elements.

\begin{itemize}
\setlength{\abovedisplayskip}{0pt}
\setlength{\itemsep}{0pt}
\setlength{\parsep}{0pt}
\setlength{\parskip}{0pt}
\item N\_\,Orbit: Integer(kind=8). The number of orbits included which should not exceed 63.
\item N\_\,Pairs: Real(kind=8). The number of pairs in the system which should not exceed the value of N\_\,Orbit. For even-mass systems, N\_\,Pairs should be an integer (N\_\,Pairs = 3 for 6 particles). For odd-mass systems, half integers are expected (N\_\,Pairs = 3.5 for 7 particles).
\item SPE: Real(kind=8),dimension(63). 1D array for the single-particle energy of each orbit, the first N\_\,Orbit elements will be used.
\item P\_\,F: Real(kind=8),dimension(63, 63). 2D array for the pairing interaction strength between the involved orbits, the first N\_\,Orbit $\times$ N\_\,Orbit elements should be initialized in a real symmetric manner.
\end{itemize}

There are four parameters that can be optionally adjusted in the source code, PairDiag.f90.

\begin{itemize}
\setlength{\itemsep}{0pt}
\setlength{\parsep}{0pt}
\setlength{\parskip}{0pt}
\item Lanc\_\,Limit: Integer(kind=8). The step size of the Lanczos algorithm, the default value is 50 and the recommended range is between 10 and 50. 
\item Lanc\_\,Error: Real(kind=8). The convergence control of the Restart Lanczos algorithm, the default value is 1$\times$$10^{-5}$, which meets general accuracy requirements.
\item Dimension\_\,Limit: Integer(kind=8). The dimension limit of the valance space, the default value is 1$\times$10$^{9}$. Truncation will be applied if the dimension exceeds the number.
\item Print\_\,Mode: Integer(kind=1). Only when the value is 0 (default), the program will print calculation information on the terminal.
\end{itemize}

\subsection{step 2: Call the Subroutine}

There is only one public subroutine that can be called in the PairDiag module.

\begin{itemize}
\setlength{\itemsep}{0pt}
\setlength{\parsep}{0pt}
\setlength{\parskip}{0pt}
\item Diag\_\,Sovler([Mode], [Block]): The subroutine calculates the pairing Hamiltonian in Eq.~\ref{hamiltonian2} or Eq.~\ref{hamiltonian3} depending on the input N\_\,Pairs. For even-mass systems, all the algorithms described above for the basis and matrix will be used. For odd-mass systems, the program calculates the $\hat{H}_{b}$ in Eq.~\ref{hamiltonian3} based on the method of even-mass systems. 
\end{itemize}

The subroutine Diag\_\,Sovler([Mode], [Block]) can optionally accept two integer parameters. Mode affects the process of Lanczos, while Blocks only affect how odd systems are handled. If the user want to use the parameter Block, the first parameter Mode must be explicitly initialized.

\begin{itemize}
\setlength{\itemsep}{0pt}
\setlength{\parsep}{0pt}
\setlength{\parskip}{0pt}
\item Mode = 0 (default): Integer(kind=4). The program performs the Restart Lanczos.
\item Mode = 1: Integer(kind=4). The program performs the Lanczos without restart.
\end{itemize}

\begin{itemize}
\setlength{\itemsep}{0pt}
\setlength{\parsep}{0pt}
\setlength{\parskip}{0pt}
\item Block = 0 (default): Integer(kind=4). For odd-mass systems, the program calculates all the possible $\hat{H}_{b}$ in Eq.~\ref{hamiltonian3}, and the solution of the subspace with the lowest ground-state energy is returned as the final result.
\item Block = integer in [1, N\_\,Orbit]: Integer(kind=4). For odd-mass systems, the program only calculates $\hat{H}_{Block}$ and return the result.
\end{itemize}

\subsection{step 3: Analyze the Outputs}

After the calculation, the information of the ground state and the Hamiltonian matrix will be stored in the following public variables. Since the details of the eigenvector are not essential and the amount of data is usually large, only the corresponding occupation numbers are saved, and also the quantity that can be derived from the occupation numbers is not given.

\begin{itemize}
\setlength{\itemsep}{0pt}
\setlength{\parsep}{0pt}
\setlength{\parskip}{0pt}
\item Energy\_\,Ground: Real(kind=8). The ground-state eigenvalue corresponding to the $\langle \phi_{g.s.}|\hat{H}|\phi_{g.s.}\rangle$.
\item Monopole\_\,Min: Real(kind=8). The minimum diagonal element of the pairing Hamiltonian matrix (i.e., the Hartree-Fock energy).
\item N\_\,Occup: Real(kind=8),dimension(63). 1D array for the occupation number of each orbit corresponding to the $2\langle \phi_{g.s.}|S^{+}_{i}S^{-}_{i}|\phi_{g.s.}\rangle$. For blocking calculation, the occupancy of the blocked orbit will be 1. 
\end{itemize}

A simple program example for the standard pairing Hamiltonian with the PairDiag module can be found in the~\ref{example}. Users can also modify the program according to their own requirements. A brief description of the variables and subroutines in the module can refer to the~\ref{brief}.

\subsection{Parallelization and Compilation}

The ``0" inversion algorithm for generating base vectors cannot be easily parallelized due to its iterative nature. The parallelization of the program is mainly performed in the process of matrix construction and diagonalization. In the present program, only OpenMP~\cite{cham2008} parallelism has been implemented in order to facilitate the implementation of the code into other nuclear structure programs. The code runs in OpenMP parallel mode by default after being compiled with the -fopenmp option in the provided Makefile. The number of parallel threads is not set by the code, so the user can set the environment variable OMP\_\,NUM\_\,THREADS to the desired number. The PairDiag code has been tested under both the ifort and gfortran compilers in the Linux system, and we recommend the ifort compiler due to the higher efficiency shown.
 
\section{Discussion}

We now briefly discuss the performance of the program in specific calculations. The reference machine is a desktop computer with an Intel Core i7-7700K 4.2GHz$\times$8 CPU and a total of 47GB memory. The compiler used is the Intel Fortran compiler (ifort version 19.0.0.117) under the Ubuntu 16.04 system. In the following calculations, single-particle energies take integers incremented from 1, and the constant pairing interaction strength $V_{ij}=G$ are used. The constant pairing interaction strength here is mainly for the convenience of comparison. In actual calculations, the matrix can take the general orbit-dependent form.

\subsection{Iteration and Convergence}

\begin{figure}
\begin{center}
\setlength{\abovecaptionskip}{0.0cm}
\includegraphics[width=0.50\textwidth]{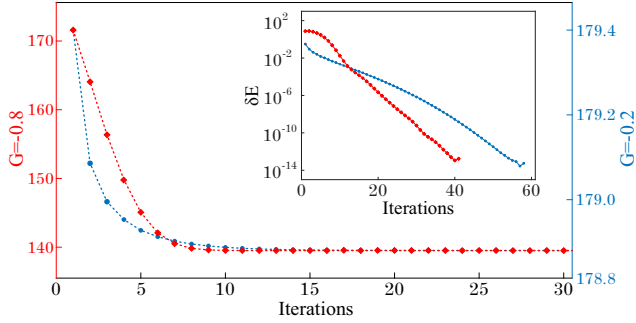}
\caption{\label{iteration} Convergence of the ground-state eigenvalue as a function of iterations with $G=-0.8$ (left red Y-axis, red curve, and red solid square dots) and $G=-0.2$ (right blue Y-axis, blue curve, and blue solid circle dots) for the space of $C_{26}^{13}$ with dimension about 10$^{7}$. The inset shows the difference between results from two consecutive iterations.}
\end{center}
\end{figure}

For the Lanczos as a projection-based approach, the larger subspace formed by more iterations often leads higher quality results. Fig.~\ref{iteration} shows the convergence of the subspace ground-state eigenvalues (Energy\_\,Ground) as a function of the number of iterations (Lanc\_\,Limit) without restart (Mode = 1). In the case where only the ground state is desired, it generally takes about 50 iterations to achieve the convergence with good accuracy. For the calculation with dimension $N$ and iteration $R$, the memory needed to store the basis and Lanczos/Ritz vectors is about 8$N(R$+2$)\times$$10^{-9}$GB in total, which means at least 41.6GB of memory is required for $N = 10^{8}$ and $R = 50$. In the program, the iteration number can be adjusted according to the local memory condition. But a larger value is recommended whenever possible. When the value of Lanc\_\,Limit is small, the accuracy of a single calculation becomes poor, and a restart strategy will be necessary.

\begin{figure}
\begin{center}
\setlength{\abovecaptionskip}{0.0cm}
\includegraphics[width=0.46\textwidth]{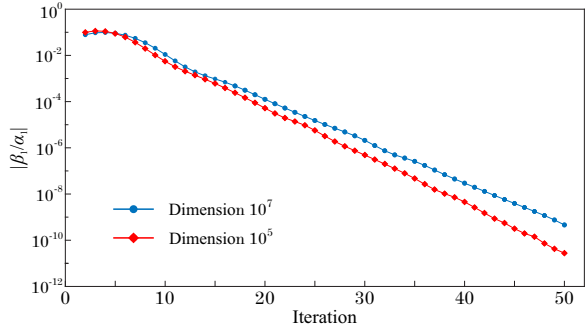}
\caption{\label{ratios} $|\,\beta_{1}/\alpha_{1}|$ after first restart as a function of iterations ($Lanc\_\,Limit$) for a system with dimension about 10$^{7}$ (blue solid circle dots) and 10$^{5}$ (red solid square dots).}
\end{center}
\end{figure}

In the restart mode (Mode = 0), the convergence condition is $|\,\beta_{i}/\alpha_{i}| \le$ Lanc\_\,Error. Since the Lanczos is restarted with the ground-state Ritz vector, the ground state will converge first with $|\,\beta_{1}/\alpha_{1}|$ reaching the threshold in a restart. $|\,\beta_{1}/\alpha_{1}|$ represents the quality of the ground-state eigenvector which can be expanded as
\begin{equation}\nonumber
\label{convergence}
\left|\frac{\beta_{1}}{\alpha_{1}}\right|=\frac{\left\|Hq_{1}-(q_{1}^{T}Hq_{1})q_{1}\right\|}{\left|q_{1}^{T}Hq_{1}\right|}=\left\|\frac{\hat{H}|q_{1}\rangle}{\langle q_{1}|\hat{H}|q_{1}\rangle}-|q_{1}\rangle\right\|
\end{equation} 
\noindent The closer $q_{1}$ is to the ground-state, the smaller the $|\,\beta_{1}/\alpha_{1}|$ will be. Fig.~\ref{ratios} shows the changes in $|\,\beta_{1}/\alpha_{1}|$ after the first restart corresponding to different Lanc\_\,Limit with $G=-0.4$. For Lanc\_\,Limit = 50, $|\,\beta_{1}/\alpha_{1}|$ after the first restart is far less than the default threshold, so the iteration will be simply terminated. When Lanc\_\,Limit is small, the Lanczos may be restarted in several rounds until convergence.

\subsection{Comparison with Other Programs}

\begin{table}
\small
\begin{center}
\setlength{\tabcolsep}{2.1mm}{
\caption{\label{lapack} Numerical comparisons of PairDiag and Lapack. $G$ are the constant pairing interaction strength, $E_{PairDiag}$ and $E_{Lapack}$ are the ground-state eigenvalues, $\Delta_{vector}$ are defined as $\sum|V^{2}_{PairDiag}(i)-V^{2}_{Lapack}(i)|$, where $V_{PairDiag}$ and $V_{Lapack}$ are the calculated ground-state eigenvectors.}
\begin{tabular}{cccc}
\hline
$G$      &  $E_{PairDiag}$        &  $E_{Lapack}$     &  $\Delta_{vector}$    \\
\hline
-0.2     &  70.106028514349  &  70.106028514349  &  8$\times$10$^{-12}$  \\
-0.4     &  66.971679958065  &  66.971679958064  &  1$\times$10$^{-10}$  \\
-0.6     &  61.289754209945  &  61.289754209946  &  8$\times$10$^{-12}$  \\
-0.8     &  53.017296011890  &  53.017296011890  &  3$\times$10$^{-13}$  \\
-1.0     &  42.931652825006  &  42.931652825006  &  4$\times$10$^{-14}$  \\
\hline
\end{tabular}}
\end{center}
\end{table}

Below we show the numerical accuracy of the PairDiag module by comparison with other programs. First, we use the Lapack package as a reference. In the space of $C_{16}^{8}$ with dimension 12870, we compared the results of the ground states between the two packages with different pairing interaction strength but fixed single-particle energies incremented from 1. In calculations, Lanc\_\,Limit was 50, and Lanc\_\,Error was $10^{-5}$. In Table~\ref{lapack}, we present the ground-state eigenvalues from the two packages with different strength $G$. For the eigenvectors in PairDiag, the user can access Q\_\,Matrix in the subroutine Result\_\,Output() described in~\ref{brief}. The negligible difference between the results indicates that the calculation of PairDiag is very reliable.

\begin{table}
\small
\begin{center}
\setlength{\tabcolsep}{4.1mm}{
\caption{\label{hspa} Ground-state eigenvalues for the standard pairing problem with G=-0.4 from the PairDiag calculation and the polynomial algorithm solution of the Richardson’s equations~\cite{guan2019}.}
\begin{tabular}{ccc}
\hline
$C_{n}^{m}$ (Dimension)              &  $E_{PairDiag}$  &  $E_{Richardson}$     \\
\hline
$C_{22}^{10}$ (6.5$\times$10$^{5}$)  &  103.0163817     &  103.0163818  \\
$C_{26}^{10}$ (5.3$\times$10$^{6}$)  &  102.2599359     &  102.2599361  \\
$C_{30}^{10}$ (3.0$\times$10$^{7}$)  &  101.5397383     &  101.5397386  \\
$C_{34}^{10}$ (1.3$\times$10$^{8}$)  &  100.8448603     &  100.8448606  \\
$C_{38}^{10}$ (4.7$\times$10$^{8}$)  &  100.1687831     &  100.1687835  \\
\hline
\end{tabular}}
\end{center}
\end{table}

Calculations with large dimension is not accessible with Lapack due to memory limitations. For the standard pairing problem, the eigenvalues can also be obtained by numerical algorithm based on the Richardson approach~\cite{rich1963}. We have developed a very efficient and robust solver for the Richardson equation \cite{qiii2015, guan2019}. In Table~\ref{hspa}, we present the ground-state eigenvalues from the PairDiag package and our Richardson solver for systems with different dimensions and with $G=-0.4$. In the calculation, Lanc\_\,Limit was set to 50 for the first three calculations, 42 and 7 for the last two due to memory limitations. As shown in Table~\ref{hspa}, the difference between the two estimates is small even for systems with large dimensions. 

\subsection{Running Time}

\begin{figure}
\begin{center}
\setlength{\abovecaptionskip}{0.0cm}
\includegraphics[width=0.46\textwidth]{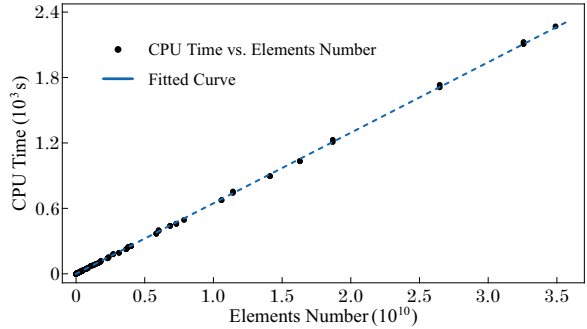}
\caption{\label{tvseh} The CPU time per Lanczos iteration as a function of the total non-zero matrix elements number with the hash search. The solid points are from the measurements and the dotted curve, $y=6.469$$\times$$10^{-8}x$, is the result of fitting.}
\end{center}
\end{figure} 

The most time-consuming part in the program is the matrix-vector multiplication during Lanczos iterations. Therefore, the running time of the entire calculation is mainly determined by the total number of iterations and the time for the matrix-vector multiplication per iteration. The total number of iterations can vary depending on interactions, spaces, and also the user's choice of error tolerance. The time cost of a single iteration is expected to be proportional to the total number of non-zero elements in the Hamiltonian matrix. For the space of $C_{n}^{m}$, the number of non-zero matrix elements is  
\begin{equation}\nonumber
\label{elements}
N = \frac{n!}{(m-1)!(n-m-1)!}
\end{equation} 
\noindent Fig.~\ref{tvseh} represents the relationship between the CPU time per iteration and the number of non-zero matrix elements from about 200 different calculations with the hash search, in which a good linearity is shown.

\begin{figure}
\begin{center}
\setlength{\abovecaptionskip}{0.0cm}
\includegraphics[width=0.46\textwidth]{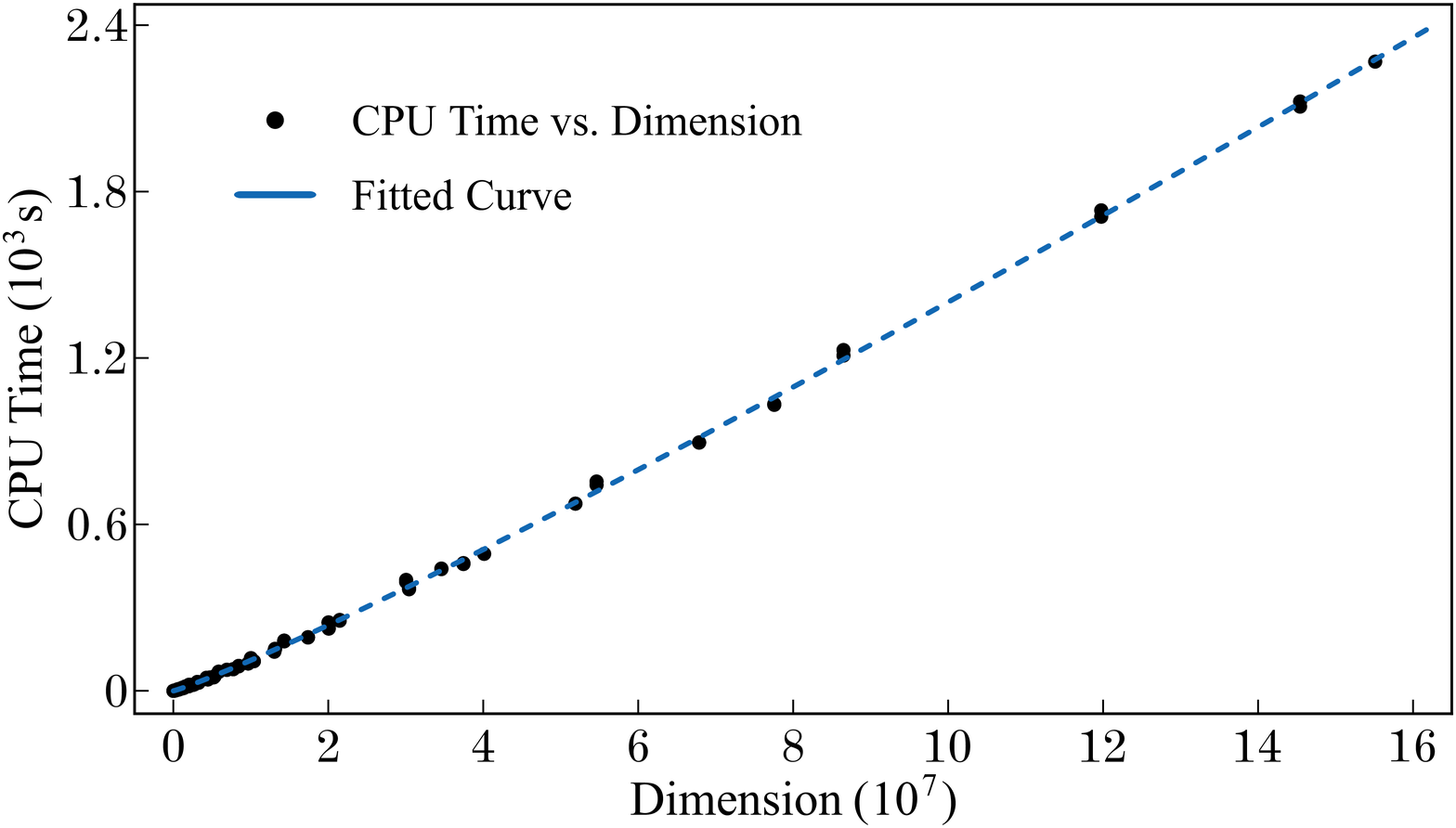}
\caption{\label{tvsdh} The CPU time per Lanczos iteration as a function of the dimension with the hash search. The dotted curve, $y=2.054$$\times$$10^{-6}x^{1.104}$, is the result of fitting.}
\end{center}
\end{figure}
 
\begin{figure}
\begin{center}
\setlength{\abovecaptionskip}{0.0cm}
\includegraphics[width=0.46\textwidth]{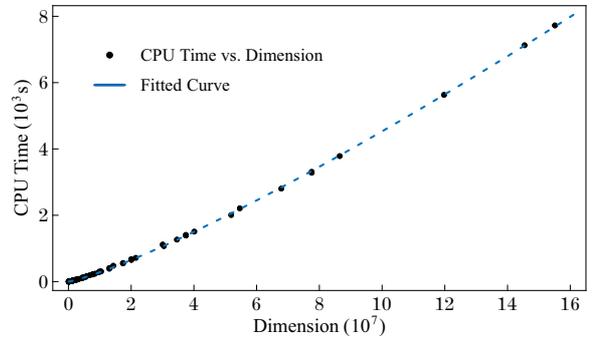}
\caption{\label{tvsdb}Same as Fig. \ref{tvsdh} but with the binary search algorithm. The dotted curve, $y=1.071$$\times$$10^{-6}x^{1.205}$, is the result of fitting.}
\end{center}
\end{figure}

Fig.~\ref{tvsdh} shows the running time of a single iteration as a function of dimension under the hash search algorithm from dimension 10$^{4}$ to 10$^{8}$. The fitting curve in the figure, $y=2.054$$\times$$10^{-6}x^{1.104}$, can be used as an empirical formula to estimate the CPU time per iteration. The CPU time is not the actual clock time. One iteration with dimension 1.5$\times$10$^{8}$ takes about 2268 seconds of CPU time, but for the case of eight cores in parallel, it only takes about 290 seconds actually, which is about one-eighth of the CPU time. The binary search is used in the program under truncations, it is more time consuming and usually takes about three times longer than the hash search as shown in Fig.~\ref{tvsdb}.

\subsection{Convergence of the truncation calculation}

\begin{figure}
\begin{center}
\setlength{\abovecaptionskip}{0.0cm}
\includegraphics[width=0.48\textwidth]{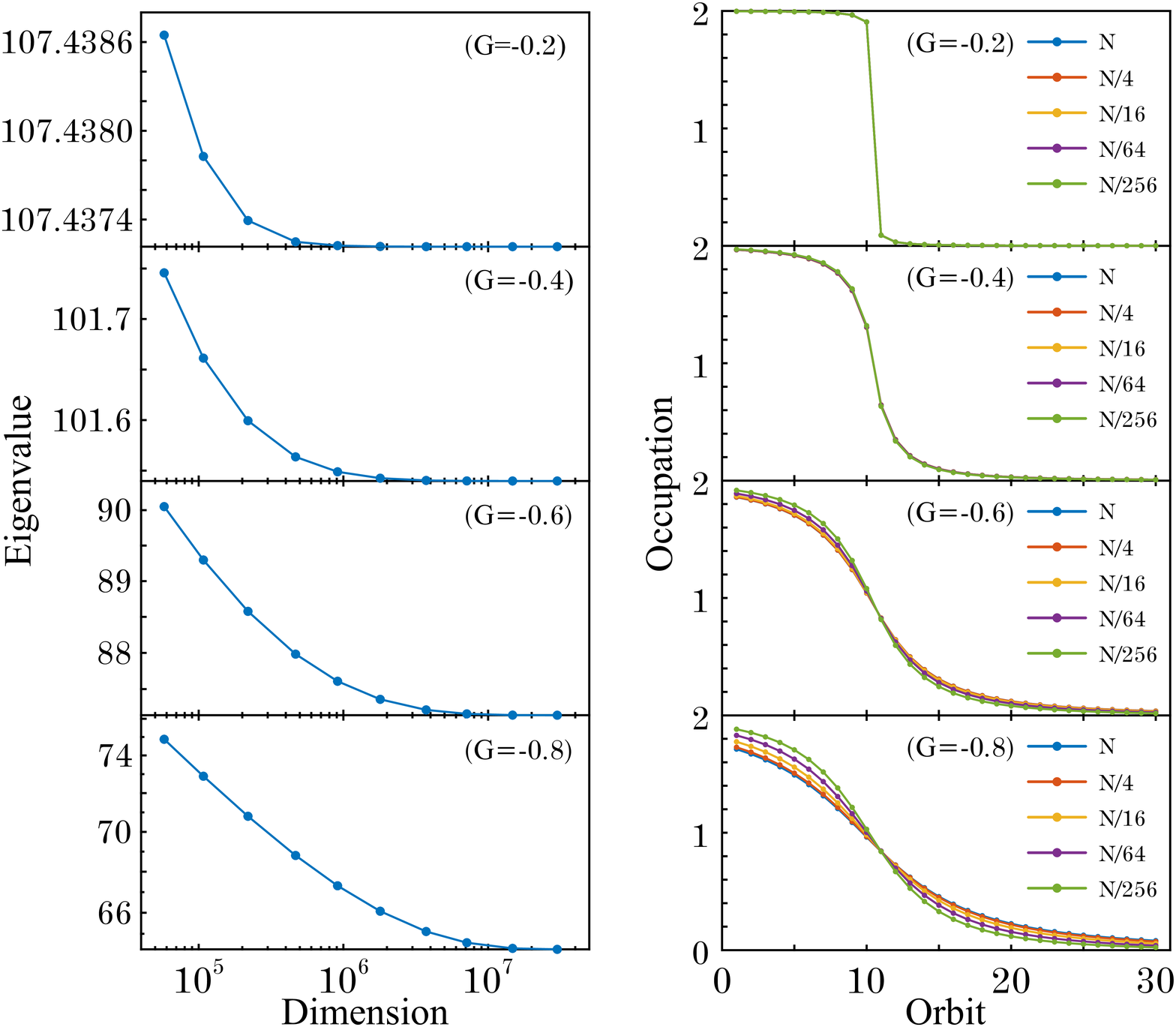}
\caption{\label{trun} Convergence of the calculation as a function of truncation level for a system with full dimension $C_{30}^{10}\sim 3.0\times10^{7}$ with different interaction strength G. Left panel shows the convergence of the ground-state energy as a function of the dimension of the truncated model space. Right panel shows the changes of occupation numbers for different orbitals under different truncations where N corresponds to the full space dimension.}
\end{center}
\end{figure}

When the dimension of the system to be calculated far exceeds the memory limit, the truncation method can be used to reduce the memory demand. The deviation of the truncated space calculation from that of the full space depends on the strength of the interaction and the level of the truncation. Fig.~\ref{trun} shows the results of the truncated calculation under different interaction strength. As can be seen from the figure, the calculation with weaker interactions can withstand more truncations. The calculated eigenvalues are more sensitive to truncation than the occupancy numbers from the corresponding eigenvectors.
 
\section{Summary}

In summary, we presented an efficient algorithm for solving the eigenproblem of the general pairing Hamiltonian based on diagonalization in the seniority-zero or seniority-one space.
We presented an efficient algorithm for solving the eigenvalue problem of the general pairing Hamiltonian based on diagonalization in the seniority-zero space. Basis vectors are generated by the `01' inversion algorithm. All the non-zero elements of the Hamiltonian matrix are evaluated on the fly. The restart Lanczos algorithm is used for the diagonalization. The present code is parallelized with OpenMP. The program can perform calculations on both the even-mass and odd-mass systems. The program also provides adjustable parameters for flexibility and a simple truncation method. The program can replace the BCS approximation in self-consistent iterative Hartree-Fock-BCS calculations. For the large space with $\Omega\sim 30$, the calculations can be done within a few hours. 

\section{Acknowledgement}

The support to X.Y. Liu provided by China Scholarship Council of No. 201700260183 during his study in Sweden is acknowledged.

\appendix

\section{A simple input example within the standard pairing Hamiltonian}\label{example}

A Fortran program for the standard paring Hamiltonian of an even-mass system using the PairDiag:

\begin{tabular}{l}
\\
\textcolor[rgb]{0,0,1}{! step 0: Load the PairDiag module}                                   \\
\textcolor[rgb]{0.7,0,0}{\textbf{use}} PairDiag                                              \\
\end{tabular}

\begin{tabular}{l}
\textcolor[rgb]{0,0,1}{! step 1: Initialize the inputs}                                      \\
\textcolor[rgb]{0.7,0,0}{\textbf{implicit none}}                                             \\
\textcolor[rgb]{0,0.5,0}{\textbf{integer}}(\textcolor[rgb]{0.7,0,0}{\textbf{kind}}=1):: i, j \\
N\_\,Orbit = \textcolor[rgb]{1,0,0.7}{10}                                                    \\
N\_\,Pairs = \textcolor[rgb]{1,0,0.7}{5}                                                     \\
\textcolor[rgb]{0.7,0,0}{\textbf{do}} i=\textcolor[rgb]{1,0,0.7}{1}, N\_\,Orbit              \\
\quad SPE(i) = i                                                                             \\
\quad \textcolor[rgb]{0.7,0,0}{\textbf{do}} j=\textcolor[rgb]{1,0,0.7}{1}, N\_\,Orbit        \\
\qquad P\_\,F(i, j) = \textcolor[rgb]{1,0,0.7}{-0.2}                                         \\
\quad \textcolor[rgb]{0.7,0,0}{\textbf{end do}}                                              \\
\textcolor[rgb]{0.7,0,0}{\textbf{end do}}                                                    \\
\end{tabular}

\begin{tabular}{l}
\textcolor[rgb]{0,0,1}{! step 2: Call the subroutine}                                        \\
\textcolor[rgb]{0.7,0,0}{\textbf{call}} Pair\_\,Diag()                                       \\
\end{tabular}

\begin{tabular}{l}
\textcolor[rgb]{0,0,1}{! step 3: Use the outputs}                                            \\
\textcolor[rgb]{0.7,0,0}{\textbf{write}}(*, *) Energy\_\,Ground                              \\
\textcolor[rgb]{0.7,0,0}{\textbf{write}}(*, *) Monopole\_\,Min                               \\
\textcolor[rgb]{0.7,0,0}{\textbf{do}} i=\textcolor[rgb]{1,0,0.7}{1}, N\_\,Orbit              \\
\quad \textcolor[rgb]{0.7,0,0}{\textbf{write}}(*, *) Occup\_\,Num(i)                         \\
\textcolor[rgb]{0.7,0,0}{\textbf{end do}}                                                    \\
\end{tabular}

\section{Brief Description of Variables and Subroutines}\label{brief}

Variables:

\begin{itemize}
\setlength{\itemsep}{0pt}
\setlength{\parsep}{0pt}
\setlength{\parskip}{0pt}
\item Lanc\_\,Limit: The step size of the Lanczos.
\item Lanc\_\,Error: In restart mode. the convergence condition in $|\,\beta_{i}/\alpha_{i}| \le$ Lanc\_\,Error.
\item Dimension\_\,Limit: Dimension limit for truncation.
\item N\_\,Orbit and N\_\,Pairs: The input number of orbits and pairs.
\item Total and Occup: The number of orbits and pairs used in the calculation.
\item B\_\,Dimension and L\_\,Dimension: The Dimension of the space and step size of the Lanczos.
\item Convergence and Truncated: Flags for convergence and truncation.
\item Run\_\,Mode, Block\_\,Mode, and Print\_\,Mode: Flags for run, block, and print.
\item Energy\_\,Blocked: Single-particle energy of the blocked orbit. 
\item Energy\_\,Ground: The output ground-state eigenvalue.
\item Monopole\_\,Min and Monopole\_\,Max: The minimum and maximum of the diagonal elements.
\item Posit\_\,Min: The position of the vector with the minimum diagonal element.
\end{itemize}

Arrays:

\begin{itemize}
\setlength{\itemsep}{0pt}
\setlength{\parsep}{0pt}
\setlength{\parskip}{0pt}
\item SPE: The 1D arrays for single-particle energies.
\item P\_\,F: The 2D array for pairing strength. 
\item B\_\,Array: The 1D array for the basis vectors. 
\item C\_\,Array: The 2D array for binomial coefficients in hash search. 
\item Q\_\,Matrix: The 2D array for the Lanczos/Ritz vectors. 
\item L\_\,Matrix: The 2D array for the Lanczos Matrix and eigenvalues. 
\item N\_\,Occup: The 1D array for occupation numbers.
\item Monopole\_\,Hist: The temporary 1D array for truncation histogram. 
\item I\_\,Vector and Q\_\,Vector: The temporary 1D array for Lanczos. 
\item O\_\,Array and V\_\,Array: The temporary 1D array for vector search.  
\item T\_\,Matrix and P\_\,Matrix: The temporary 2D array for QR.  
\end{itemize}

Subroutines and functions:

\begin{itemize}
\setlength{\itemsep}{0pt}
\setlength{\parsep}{0pt}
\setlength{\parskip}{0pt}
\item Diag\_\,Solver(Mode, Block): The only public subroutine that analyzes the input.
\item Combin\_\,Num(N, M): The function retruns the value of $C_{N}^{M}$ according to Eq.~\ref{extcmn}.
\item Next\_\,State(State): The subroutine operates the input (State) according to the `01' inversion algorithm.~\ref{conversion}.
\item Monopole\_\,E(State): The function retruns the diagonal element value of input (State) according to Eq.~\ref{diagonal}.
\item Trun\_\,State(State): The function retruns logical .TURE. if Monopole\_\,E(State) $\le$ Monopole\_\,Trun.
\item Vector\_\,Initialize() and Vector\_\,Restart(): The subroutines that initialize the starting vector of Lanczos to $[1,0,\cdots,0]^{T}$ and Q\_\,Matrix(1, :).
\item Bina\_\,State(D, L) and Hash\_\,State(D, L): The subroutines that calculate non-zero matrix elements and positions related to the input State using binary and hash search.
\item Lanczos\_\,Iteration(): The subroutine for Lanczos iteration from starting vector I\_\,Vector.
\item QR\_\,Decompose(): The subroutine for QR decompose to the L\_\,Matrix.
\item Even\_\,System() and Odds\_\,System(): The subroutine for the even-mass and the odd-mass system  calculation.
\item Lanczos\_\,Restart(): The subroutine that combines the Lanczos\_\,Iteration() and QR\_\,Decompose() in Restart Mode.
\item Initialize(): The subroutine that allocates memory for dynamic arrays and initializes basis vectors.
\item Results\_\,Output(): The subroutine that calculates the occupation numbers and other outputs.
\item Destory(): The subroutine that releases all dynamic memories.
\end{itemize}

\end{document}